\documentclass[aps,prl,preprint,superscriptaddress]{revtex4-2}

\usepackage{graphicx}
\usepackage[T1]{fontenc}

\begin{document}


\title{Role of topology in scaling laws for studying mechanics in open-porous solids: Moving beyond classical Gibson-Ashby scaling}


\author{Ameya Rege}
\email{ameya.rege@utwente.nl}
\affiliation{%
 University of Twente, Department of Mechanics of Solids, Surfaces \& Systems, P.O. Box 217, 7500 AE, Enschede, The Netherlands
}%


\date{\today}

\begin{abstract}
The elastic modulus of porous materials is commonly described using power-law scaling relations with relative density, where the scaling exponent is often interpreted in terms of the underlying deformation mechanism. However, in highly disordered porous networks, changes in density are generally accompanied by changes in network topology, which can substantially modify the apparent scaling behavior. In this paper, we propose a topology-informed framework that separates the intrinsic mechanical contribution from the effects of network structure. Three representative topological descriptors are considered: the mean coordination number, the fraction of the load-bearing backbone, and the tortuosity of the load paths. For each case, the corresponding density-dependent contribution to the apparent modulus-scaling exponent is derived and analyzed. The results show that variations in connectivity, mechanical participation of the solid phase, and load-path efficiency can all lead to apparent scaling exponents exceeding the intrinsic exponent associated with the local deformation mechanism. These effects are particularly pronounced at low relative densities, where network topology evolves most strongly. The framework provides a physically interpretable basis for understanding anomalous modulus-density scaling in disordered porous materials and highlights the need to consider topology explicitly alongside relative density.
\end{abstract}


\maketitle

Open-porous solids often exhibit a scaling behavior for their mechanical, thermal, or acoustic properties with respect to their density \cite{ashby1997cellular}. In this paper, we focus on the classical scaling relation of the Young modulus versus density. A power law scaling is used as,

\begin{equation}
    \bar{E} \sim C\bar{\rho}^{m_0},
\end{equation}

\noindent where $\bar{(\cdot)}$ denotes the apparent property of the material. Therefore, $\bar{E} = \frac{E}{E_s}$ and $\bar{\rho} = \frac{\rho}{\rho_s}$. The quantity $(\cdot)_s$ denotes the skeletal property. The scaling exponent $m_0$ for an ideal foam-like structure is 2. This quadratic relation was well established by Gibson and Asbhy by considering a bending-dominated local deformation mode of the cell walls/struts \cite{gibson1982mechanics}. This $m_0$ tends to $ 1$ for a stretch dominated behavior. Such an axial-mode of deformation was considered as the primary mechanism of deformation in porous materials in one of the first reports by Gent and Thomas \cite{gent1959deformation}. Therefore, for any open-porous solid, the scaling law should move between 1-2 purely considering the mechanics of the struts. However, several open-porous materials often show scaling exponents $>2$, e.g., porous graphene with 2.7 \cite{qin2017mechanics}, silica aerogels show 3.6 \cite{aney2025origin}, while some architected materials show 5 \cite{cuan2020compressive}. For the case of aerogels, Ma. et al. \cite{ma2000mechanical} ruled out the role of dangling mass in describing the exponent while recently Aney et al. \cite{aney2025origin} ruled out the effect of the strut morphology. It has been pointed out that the topology of the network must play a critical role in dictating this property \cite{huber2018connections, topolnicki2026direction}. This is because the stiffness of a material must not only be controlled by how much solid is present but also by how that solid is connected. As a thought experiment, one may consider two networks having the same apparent density $(\bar{\rho})$ but: (a) one may have many well-connected junctions, while (b) the other may have dead ends and poorly connected struts. The way the load gets transmitted will be very different in each case. However, to the best of the knowledge of the author, there are not many studies quantifying this process by means of mathematical models that describe how the exponent moves beyond the classical Gibson-Ashby quadratic scaling.  A closely related study on nanoporous gold demonstrated that macroscopic stiffness and strength are highly sensitive to network topology, with topology-dependent modifications required to classical density-based scaling laws \cite{mangipudi2016topology}. Similarly, Sohn et al. demonstrated that the effective Young's modulus of random network nanomaterials depends not only on solid fraction but also on topological genus, supporting a separable topology-density description of elasticity \cite{sohn2024scaling}.

In this paper, we explore a few topological descriptors that map out the scaling law beyond 2. The global idea is to separate the effect of topology and the effect of mechanics on the scaling relation. In order to maintain the scaling with density, one must describe the topological descriptor as a function of density. Therefore, one may propose a rewriting of Eq. 1 as 

\begin{equation}
    \bar{E} = C(\bar{\rho}) \bar{\rho}^{m_0},
\end{equation}

\noindent where $C(\bar{\rho})$ is the topological descriptor acting as a prefactor and is dependent on the density. The second part in this multiplicative decomposition of the scaling relation is the one controlled by local mechanics of struts, being bending or stretching dominated. The question now appears, how may one describe these topological descriptors. In the following, we propose three such descriptors. 

\section{Case 1: Mean coordination number}

Let us consider mapping the network of a porous solid as a graph. For a graph, given $N_v$ vertices and $N_e$ edges, the mean coordination number $z$ is given by

\begin{equation}
    z = \frac{1}{N_v}\sum_{i=1}^{N_v}k_i,
\end{equation}

\noindent where $k_i$ is the degree of node $i$. Using the identity $\sum_ik_i = 2N_{e}$, one gets

\begin{equation}
    z = \frac{2N_e}{N_v}. 
\end{equation}

The coordination number therefore represents the average number of struts connected to a node. So at a fixed density, larger $z$ should generally imply larger $E$. So we replace the prefactor $C$ from Eq. 1 to a function of $z$, say $C(z)$ as

\begin{equation}
    \bar{E}\sim \bar{\rho}^{m_0}C(z).
\end{equation}

\noindent From past experience, we know that there is usually some critical connectivity below which the network is not mechanically robust/stable. Therefore, we introduce a critical connetivity $z_c$ into the equation as

\begin{equation}
    \bar{E}\sim \bar{\rho}^{m_0}C(z-z_c),
\end{equation}

\noindent where we call $z_c$ as the threshold coordination for effective load transfer. As $z\to z_c$, the network becomes unstable while as $z > z_c$ the stiffness rises. Therefore, $z-z_c$ could be thought of as a topological measure of how far the network is from marginal rigidity. The simplest way of defining $C(z-z_c)$ is

\begin{equation}
    \bar{E} = C_0\bar{\rho}^{m_0}(z-z_c)^{m_1}.
\end{equation}

Now we introduce a crucial element. The coordination number $z$ must be dependent on $\bar{\rho}$. Therefore

\begin{equation}
    z = z(\bar{\rho}).
\end{equation}

\noindent Therefore, Eq. 7 becomes

\begin{equation}
    \bar{E} = C_0\bar{\rho}^{m_0}(z(\bar{\rho})-z_c)^{m_1}.
\end{equation}

\noindent Taking logarithms,

\begin{equation}
    \ln \bar{E} = \ln C_0 + m_0\ln \bar{\rho} + m_1\ln (z(\bar{\rho})-z_c),
\end{equation}

\noindent and now differentiating with respect to $\ln \bar{\rho}$ yields

\begin{equation}
    \frac{d\ln \bar{E}}{d\ln \bar{\rho}} = m_0 + m_1\frac{d\ln (z-z_c)}{d\ln \bar{\rho}}.
\end{equation}

\noindent Therefore, the apparent scaling exponent $m$ becomes

\begin{equation}
    m = m_0 + m_1\frac{d\ln (z-z_c)}{d\ln \bar{\rho}},
\end{equation}

\noindent where $m_0$ depends on the mechanics of the local struts, e.g., $m_0\sim 2$ for a bending dominated deformation mode. $m_1$ describes how strongly the stiffness depends on connectivity. Lastly, the term $\frac{d\ln (z-z_c)}{d\ln \bar{\rho}}$ measures how the connectivity changes when one changes the density. Eq. 12 also works for the case that topology does not change with density. For instance $z=$ constant results in $m=m_0$, which aligns with the classical Gibson-Ashby scaling. As connectivity decreases with decreasing density, which is typically the case for e.g., aerogels, we have the $m>m_0$, as observed in experiments. Likewise, increase in connectivity with density also results in the same behavior. $m_1$ can also be thought of as an amplification factor in relation to topological sensitivity. So $\frac{d\ln (z-z_c)}{d\ln \bar{\rho}}$ quantifies how connectivity evolves with density, whereas $m_1$ quantifies how strongly this topological evolution influences the effective stiffness scaling.

\begin{figure*}[htbp]
    \centering
    \includegraphics[width=\textwidth]{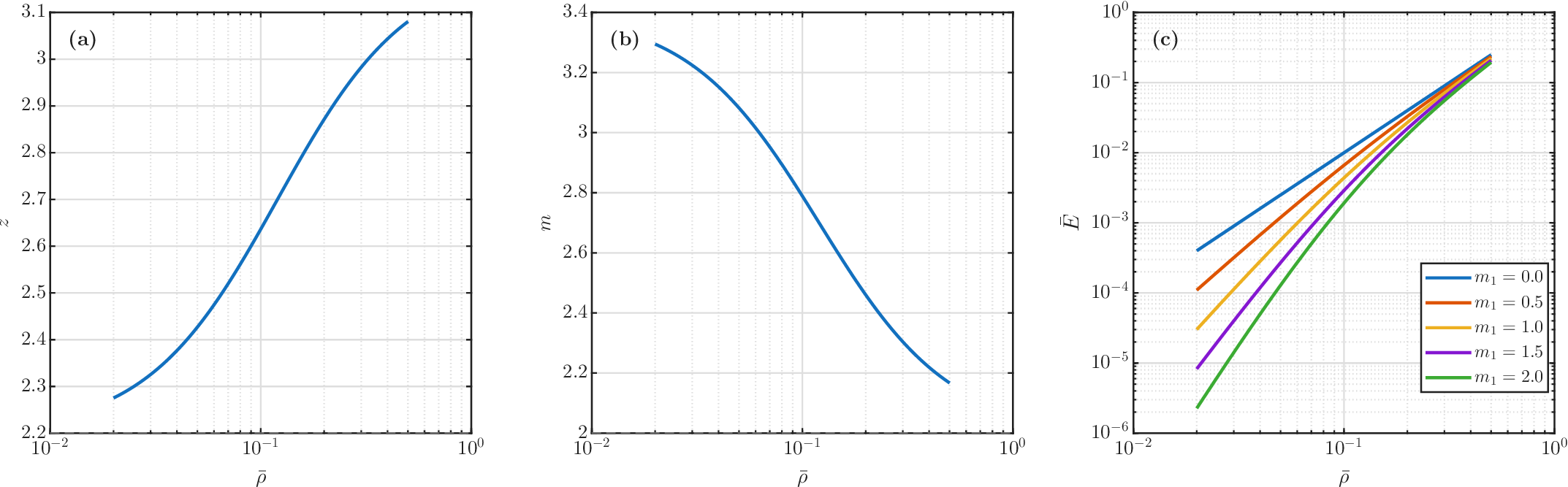}
    \caption{Effect of mean coordination number as a topological descriptor in Eq. 2. (a) Relation between the mean coordination number and the apparent density, (b) for a constant $m_0=2$, the effect of apparent density on $m$, as in Eq. 12, and (c) the effect of $m_1$ on the modulus-scaling relation for a constant $m_0=2$.}
    \label{fig:coordination}
\end{figure*}

Figure~\ref{fig:coordination}a shows an exemplary evolution of the mean coordination number $z$ with relative density $\bar{\rho}$. The coordination number increases with density and approaches a saturation value at larger $\bar{\rho}$, whereas at low densities it approaches the critical coordination $z_c$. This low-density regime is particularly important because the network lies close to marginal rigidity, such that even relatively small changes in connectivity can substantially alter the ability of the structure to transfer load. Conversely, once the coordination number begins to saturate, further increases in density result in progressively smaller topological changes.

The consequence of this density-dependent connectivity for the apparent scaling exponent is shown in Fig.~\ref{fig:coordination}b. At low relative densities, the strong variation of $z-z_c$ with $\bar{\rho}$ produces a substantial topological contribution, resulting in an apparent exponent $m$ larger than the intrinsic exponent $m_0$. With increasing density, the coordination number becomes progressively less sensitive to $\bar{\rho}$, and $m$ approaches $m_0$. Thus, the classical density scaling is recovered when the network topology becomes approximately independent of density, a case which could for instance occur in high-density foams. Importantly, this shows that an experimentally measured scaling exponent exceeding the value expected for the underlying strut-level deformation mechanism does not necessarily indicate a change from, for example, bending- to another deformation mode. Instead, part of the apparent exponent can originate from the simultaneous evolution of network connectivity with density.

This effect is further illustrated in Fig.~\ref{fig:coordination}c, where the modulus-density relationship is shown for different values of the topology-sensitivity exponent $m_1$, while $m_0$ is kept fixed. For $m_1=0$, the coordination number does not contribute to the stiffness scaling and the classical dependence governed by $m_0$ is recovered. Increasing $m_1$ progressively amplifies the influence of connectivity on the effective modulus. The separation between the curves is particularly pronounced at low densities, where the network is closest to the connectivity threshold, whereas the influence becomes weaker at higher densities as $z$ saturates. Figure~\ref{fig:coordination}c therefore demonstrates that porous solids with the same relative density and the same intrinsic deformation exponent $m_0$ can nevertheless exhibit markedly different modulus-density scaling depending on the sensitivity of their mechanical response to connectivity. In this sense, $m_1$ provides an independent measure of the mechanical importance of network coordination.

\section{Case 2: Fraction of lead-bearing backbone}

In colloidal solids, e.g., in carbon black aggregates or porous materials like silica aerogels, it is well known that not all of the solid in the porous material bears the applied load. Therefore, there is a fraction of the network, the so called backbone or load-bearing backbone, while there is the remaining fraction that is mechanically inefficient. The latter could compose of dangling ends, weakly connected clusters, etc. Therefore, one can think of the backbone fraction as a way to distinguish between total apparent density $\bar{\rho}$ and the mechanically active relative density. Let us break this down mathematically.

\begin{figure*}[htbp]
    \centering
    \includegraphics[width=\textwidth]{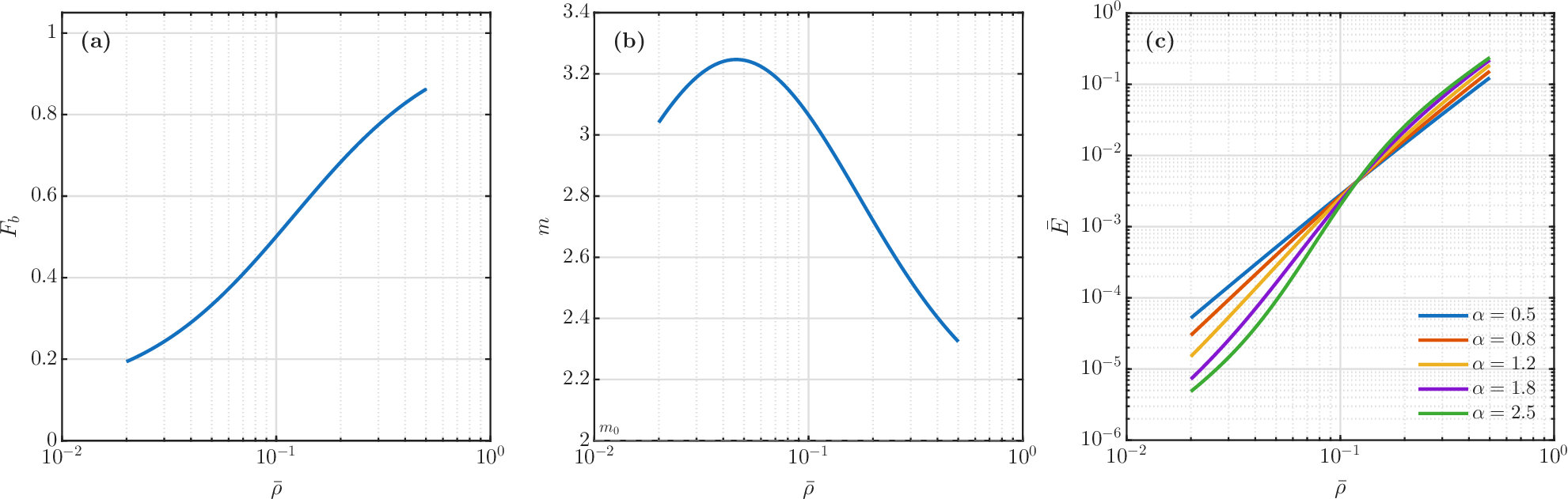}
    \caption{Effect of backbone fraction as a topological descriptor in Eq. 2. (a) Relation between the backbone fraction and the apparent density, (b) for a constant $m_0=2$, the effect of apparent density on $m$, as in Eq. 19, and (c) the effect of $\alpha$ on the modulus-scaling relation for a constant $m_0=2$.}
    \label{fig:backbone}
\end{figure*}

Let the total mass and volume of the porous network be $M$ and $V$. Now let the part that belongs to the load-bearing backbone have $M_b$ and $V_b$ as counterparts. We now define the backbone fraction as

\begin{equation}
    F_b(\bar{\rho}) = \frac{M_b}{M} = \frac{V_b}{V},
\end{equation}

\noindent where $0<F_b<1$. For the case $F_b=1$, all of the network is mechanically active. For the cases $F_b<1$ fraction of the solid in the network does not bear the applied load. So now we define the effective amount of solid network participating in bearing the load as

\begin{equation}
    \bar{\rho}_b = F_b(\bar{\rho})\bar{\rho},
\end{equation}

\noindent where $\bar{\rho}_b$ is the backbone apparent density. Now coming back to our classical scaling relation of $\bar{E}$, one can say that if only $F_b$ is actually the one bearing the load, the elastic modulus must be controlled by the backbone apparent density $\bar{\rho}_b$ and not $\bar{\rho}$. Therefore, one can write

\begin{equation}
    \bar{E}\sim C\bar{\rho}_b^{m_0}.
\end{equation}

\noindent and from Eq. 14, one can write

\begin{equation}
    \bar{E} = C\bar{\rho}^{m_0}F_b(\bar{\rho})^{m_0}.
\end{equation}

\noindent While one notices that we still have exponent only as $m_0$, if we take a logarithm of this equation

\begin{equation}
    \ln \bar{E} = \ln C + m_0\ln \bar{\rho} + m_0 \ln F_b(\bar{\rho}),
\end{equation}

\noindent and similar to Case 1, differentiating this now with $\ln \rho$, we get

\begin{equation}
    \frac{d\ln \bar{E}}{d \ln \bar{\rho}} = m_0 + m_0\frac{d\ln F_b}{d\ln \bar{\rho}}.
\end{equation}

\noindent Therefore, the apparent exponent becomes

\begin{equation}
    m = m_0\left(1+\frac{d\ln F_b}{d\ln \bar{\rho}}\right). 
\end{equation}

\noindent One can observe that if $F_b$ is constant, we get $m=m_0$, corresponding to the classical Gibson-Ashby scaling. But if $F_b$ increases with density, we have $m>m_0$.

The evolution of the load-bearing backbone fraction with relative density is shown in Fig.~\ref{fig:backbone}a. At low relative densities, only a limited fraction of the solid network belongs to mechanically continuous load-bearing pathways. The remaining solid may be associated with dangling branches, weakly connected clusters, or other structural elements that contribute to the total apparent density without participating efficiently in load transfer. As $\bar{\rho}$ increases, $F_b$ increases and progressively approaches unity, indicating that an increasing fraction of the available solid becomes incorporated into the mechanically active backbone. Figure~\ref{fig:backbone}a therefore illustrates the distinction between the total amount of solid contained in the porous network and the fraction of this solid that effectively contributes to its macroscopic stiffness.

\begin{figure*}[htbp]
    \centering
    \includegraphics[width=\textwidth]{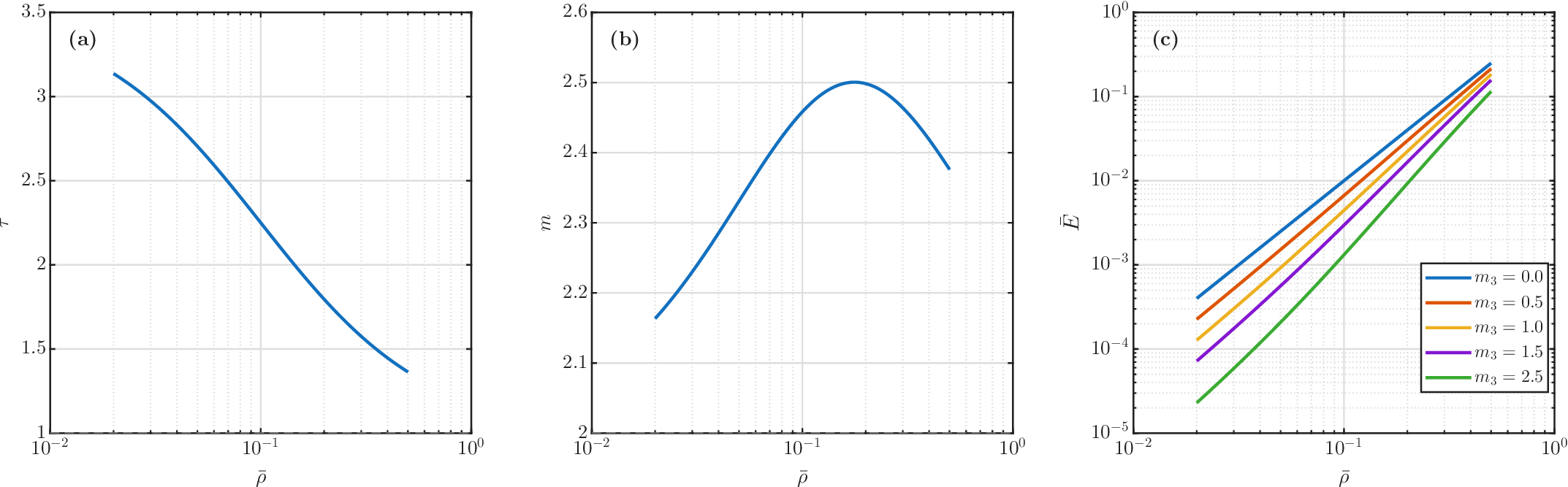}
    \caption{Effect of tortuosity as a topological descriptor in Eq. 2. (a) Relation between the tortuosity and the apparent density, (b) for a constant $m_0=2$, the effect of apparent density on $m$, as in Eq. 22, and (c) the effect of $m_3$ on the modulus-scaling relation for a constant $m_0=2$.}
    \label{fig:tortuosity}
\end{figure*}

The consequence of this evolving load-bearing fraction for the apparent density-scaling exponent is shown in Fig.~\ref{fig:backbone}b. In the low-density regime, $F_b$ changes strongly with $\bar{\rho}$, such that the development of the mechanically active backbone provides an additional contribution to the modulus-density scaling. The apparent exponent consequently exceeds the intrinsic value $m_0$. With increasing relative density, the backbone fraction progressively saturates and its relative change with density becomes smaller. Accordingly, the apparent exponent approaches $m_0$. The classical scaling is therefore recovered when the fraction of mechanically active material becomes approximately independent of density. This result has a direct physical interpretation. Increasing the relative density of a sparse disordered network does not merely introduce additional solid material. It can simultaneously convert previously inactive or poorly connected regions into continuous load-bearing pathways. The resulting increase in stiffness therefore reflects two concurrent effects: an increase in the total solid fraction and an increase in the fraction of that solid that participates mechanically. Consequently, an experimentally measured exponent larger than $m_0$ does not necessarily imply a different local deformation mechanism. Instead, it may partly result from the progressive formation of the load-bearing backbone as the network density increases. The sensitivity of this behavior to the evolution of network topology is demonstrated in Fig.~\ref{fig:backbone}c. Here, $m_0$ is held constant for all curves, while the parameter $\alpha$, which controls how rapidly $F_b$ evolves with relative density, is varied. Changing $\alpha$ alters the density range over which the mechanically active backbone develops and consequently modifies the shape of the modulus-density relationship. Larger values of $\alpha$ produce a more abrupt development of the load-bearing backbone around the characteristic density, resulting in a correspondingly stronger change in the local slope of the modulus curve. Smaller values of $\alpha$, in contrast, distribute this topological evolution over a broader density range and lead to a smoother transition in the stiffness scaling. Importantly, the differences between the curves in Fig.~\ref{fig:backbone}c arise despite the fact that the intrinsic mechanical exponent $m_0$ remains unchanged. 

The variation in the apparent modulus scaling therefore originates entirely from differences in how the load-bearing topology evolves with density. This demonstrates a central feature of the present framework: changes in experimentally observed scaling behavior can arise from the density dependence of network topology without requiring any modification of the underlying strut-level deformation mechanism. The backbone description also captures an aspect of topology that is distinct from the mean coordination number considered in Case~1. Coordination characterizes the average connectivity of the nodes, whereas $F_b$ quantifies the fraction of the complete solid network that is incorporated into mechanically continuous load-bearing pathways. A network may therefore exhibit a finite or even relatively large mean coordination while still containing a considerable amount of mechanically inactive material. The backbone fraction consequently provides a complementary measure of topological efficiency, particularly for highly disordered networks such as colloidal aggregates and aerogels.

\section{Case 3: Tortuosity}

Tortuosity measures how convoluted (longer) the load path is compared to a straight line. For a solid backbone,

\begin{equation}
    \tau = \frac{l_{\mathrm{path}}}{l_{\mathrm{straight}}} \geq 1,
\end{equation}

\noindent where $l_{\mathrm{path}}$ is the actual path along the backbone between two points, while $l_{\mathrm{straight}}$ is the shortest distance between the two points. Therefore, $\tau=1$ demonstrates perfectly straight paths, like lattice structures, and $\tau>1$ shows wavy indirect load paths. For the case of colloids or aerogels, $\tau$ can be significantly larger than 1. Therefore, in some sense, tortuosity introduces geometric inefficiency in load transfer. We now incorporate this effect as a prefactor in the scaling relation. Let us introduce the tortuosity as a penalty in Eq. 1 as

\begin{equation}
    \bar{E} = C\bar{\rho}^{m_0}\tau(\bar{\rho})^{-m_3},
\end{equation}

\noindent where $m_3>0$ demonstrates sensitivity to tortuosity. One may also notice that the exponent is introduced as negative. This is because higher tortuosity leads to lower stiffness. Taking the logarithm and following the procedure in the first two cases, gives us

\begin{equation}
    m = m_0 - m_3\frac{d\ln \tau}{d\ln \bar{\rho}}.
\end{equation}

\noindent As density decreases, the network becomes more open and disordered leading to the paths becoming more say winding. Therefore tortuosity increases. This implies that $\frac{d\ln \tau}{d\ln \bar{\rho}}$ becomes negative, and therefore, $m>m_0$. 

Figure~\ref{fig:tortuosity}a shows the assumed evolution of tortuosity with relative density. In contrast to coordination number and backbone fraction, for which increasing values generally represent improved load transfer, tortuosity acts as a measure of geometrical inefficiency. Sparse networks exhibit larger values of $\tau$, corresponding to longer and more convoluted load paths, whereas increasing density leads to progressively straighter and more direct pathways. At sufficiently high densities, the tortuosity approaches its lower limiting value, indicating that further densification produces increasingly small changes in load-path geometry.

The resulting influence on the apparent scaling exponent is shown in Fig.~\ref{fig:tortuosity}b. Since tortuosity decreases as the network becomes denser, the improvement in load-path efficiency contributes in addition to the direct effect of increasing solid fraction. The apparent exponent therefore exceeds $m_0$ in the density range over which tortuosity evolves strongly. As the load paths become progressively straighter and $\tau$ approaches its limiting value, this additional topological contribution decreases and $m$ converges toward $m_0$.

Figure~\ref{fig:tortuosity}c illustrates how the tortuosity-sensitivity exponent $m_3$ modifies the modulus-density relationship while $m_0$ remains fixed. For $m_3=0$, tortuosity has no influence and the stiffness follows the intrinsic density scaling. Increasing $m_3$ progressively penalizes highly tortuous networks, resulting in substantially lower stiffness at small $\bar{\rho}$. As density increases and the load paths straighten, the difference between the curves decreases. Thus, $m_3$ determines how strongly geometrical inefficiency in the load path is translated into a macroscopic stiffness reduction.

The physical interpretation of Fig.~\ref{fig:tortuosity} is complementary to those of Figs.~\ref{fig:coordination} and~\ref{fig:backbone}. Coordination number describes whether a sufficiently connected network exists, while the backbone fraction describes how much of the solid belongs to that connected load-bearing network. Tortuosity, in contrast, describes how efficiently the load is transmitted through those pathways once they exist. A porous network can therefore possess a well-developed load-bearing backbone and sufficient connectivity while still exhibiting low stiffness because the force-transmission paths are highly convoluted. The three descriptors consequently capture distinct, although potentially coupled, aspects of network topology. The trends in Fig.~\ref{fig:tortuosity} show that changes in load-path geometry alone can generate an apparent density exponent exceeding that associated with the underlying local deformation mechanism.

In summary, Figs.~\ref{fig:coordination}-\ref{fig:tortuosity} demonstrate that the apparent modulus-density scaling of a porous material cannot necessarily be interpreted solely in terms of the local deformation mechanism. The intrinsic exponent $m_0$ characterizes the mechanics of the solid skeleton, whereas the experimentally observed exponent $m$ may additionally reflect systematic changes in topology as the density varies. The three cases considered here separate this topological contribution into connectivity through the coordination number, mechanical participation through the load-bearing backbone fraction, and load-path efficiency through tortuosity. In all three cases, the strongest deviations from the intrinsic scaling occur in the low-density regime, where network topology changes most rapidly and where highly porous materials are generally closest to marginal mechanical stability. The framework therefore offers a physical interpretation for scaling exponents substantially larger than classical Gibson-Ashby expectations without requiring a corresponding change in the underlying strut-level deformation mode. More generally, the results suggest that density should not be regarded as an independent structural variable, namely, in disordered porous solids, changing density also changes the topology of the network. Explicitly accounting for this coupling provides a more physically interpretable basis for understanding and comparing stiffness scaling across different classes of porous materials.

\bibliography{apssamp}

\end{document}